\documentclass[prd,preprintnumbers,eqsecnum,floatfix,letterpaper,superscriptaddress,nofootinbib]{revtex4}

\usepackage{amsfonts,amsmath,units,wasysym,epsfig,graphicx,verbatim,color,subfigure,graphicx}
\usepackage{amsmath}
\usepackage{amssymb}
\usepackage{amsfonts}
\usepackage{mathrsfs}
\usepackage{bm}
\usepackage{color}

\def\be{\begin{equation}}
\def\ee{\end{equation}}
\def\bea{\begin{eqnarray}}
\def\eea{\end{eqnarray}}

\def\lsim{\mathrel{\rlap{\lower4pt\hbox{\hskip1pt$\sim$}}
    \raise1pt\hbox{$<$}}}                % less than or approx. symbol
\def\gsim{\mathrel{\rlap{\lower4pt\hbox{\hskip1pt$\sim$}}
    \raise1pt\hbox{$>$}}}                % greater than or approx. symbol

\begin{document}

\newcommand{\rhat}{\hat{r}}
\newcommand{\iotahat}{\hat{\iota}}
\newcommand{\phihat}{\hat{\phi}}

\newcommand{\WSU}{Department of Physics \& Astronomy, Washington State University,
1245 Webster, Pullman, WA 99164-2814, U.S.A}

\newcommand{\IUCAA}{Inter-University Centre for Astronomy and
  Astrophysics, Post Bag 4, Ganeshkhind, Pune 411 007, India}

\renewcommand{\thesection}{\Roman{section}} 
\renewcommand\thesubsection{\Alph{subsection}}

\title{Multi-detector null-stream-based $\chi^2$ statistic for
  compact binary coalescence searches}

\author{William Dupree}
\email{william.dupree@wsu.edu}
\affiliation{\WSU}

\author{Sukanta Bose}
\email{sukanta@iucaa.in}
\affiliation{\WSU}
\affiliation{\IUCAA}

\date{\today}

\begin{abstract}

We develop a new multi-detector signal-based discriminator to improve
the sensitivity of searches for gravitational waves from compact
binary coalescences. The new statistic is the traditional $\chi^2$
computed on a null-stream synthesized from the gravitational-wave detector
strain time-series of three detectors. This null-stream-$\chi^2$ statistic can be extended
to networks involving more than three detectors as well.
The null-stream itself was proposed as a discriminator
between correlated unmodeled signals in multiple detectors, such as arising from
a common astrophysical source, and uncorrelated noise transients. It
can be useful even when the signal model is known, such as for compact
binary coalescences. The traditional
$\chi^2$, on the other hand, is an effective discriminator when the
signal model is known and lends itself to the matched-filtering
technique. The latter weakens in its effectiveness when a signal
lacks enough cycles in band; this can happen for high-mass black hole
binaries. The former weakens when there are concurrent noise
transients in different detectors in the network or the detector
sensitivities are substantially different. Using simulated binary
black hole signals, noise transients and strain for
Advanced LIGO (in Livingston and Hanford) and Advanced Virgo detectors,
we compare the performance of the null-stream-$\chi^2$ statistic with that
of the traditional $\chi^2$ statistic using receiver-operating characteristics. 
The new statistic may form the basis for better
signal-noise discriminators in multi-detector searches in the future.
\end{abstract}

\preprint{[LIGO-18000334]}

\maketitle

\section{Introduction} \label{introduction}

The last few years have witnessed major progress in gravitational wave (GW)
astronomy~\cite{Abbott:2016blz,Abbott:2016nmj,Abbott:2017vtc,Abbott:2017oio,TheLIGOScientific:2017qsa},
with the LIGO and Virgo detectors~\cite{ligo,virgo} successfully observing
numerous black hole-black hole (BBH) mergers as well as one neutron
star-neutron star merger (BNS)~\cite{LIGOScientific:2018mvr} --
jointly called compact binary coalescences (CBCs).
This progress has led to the growth
of ground-based detection efforts, in signal processing as well as the
planning of new detectors and sites. As detections become more common
there is a growing need for statistical analysis that improves our
ability to
separate signals from spurious noise transients~\cite{TheLIGOScientific:2017lwt,Bose:2016sqv,Bose:2016jeo,Mukund:2016thr,Zevin:2016qwy,Nuttall:2018xhi,Berger:2018ckp,Cavaglia:2018xjq,Walker:2017zfa}. In this study we
focus on this need, but with an emphasis on utilizing three or more detectors' data streams in unison with a network-wide statistic.
There have been methods proposed in coherent
searches~\cite{Bose:1999pj,Pai:2000zt,Bose:2011km,Harry:2010fr,Talukder:2013ioa}
in data from multiple detectors 
for improving the separation of false positives from
GW signals. These coherent methods involve multiple statistical tests,
some applied separately to individual detectors while others applied to the network data jointly. Two such methods are the $\chi^2$-distributed statistics~\cite{Allen:2004gu,Babak:2012zx,Dhurandhar:2017aan}
and the network null-stream~\cite{Guersel:1989th}.

It is well known how Gaussian random noise influences the construction
of GW search statistics for modeled signals, and how $\chi^2$
distributed statistics can distinguish between signals and certain
types of non-Gaussian noise transients, or ``glitches" that are
sometimes present in the detector
data~\cite{Dhurandhar:2017aan}. These $\chi^2$ tests can take several
forms (since the sum of squared normally distributed random variables
is $\chi^2$ distributed), but one of the more common tests suited to
distinguishing glitches from signals is described in
Allen~\cite{Allen:2004gu}, which relies on dividing the matched
filter~\cite{Helstrom} over a putative signal's band into several
sub-bands and checking for consistency between the distribution of its
anticipated and observed values in them. On the other hand, for separating signals from glitches in a network of detectors the null stream~\cite{Guersel:1989th,Chatterji:2006nh,Klimenko:2008fu} has been found of some use since it is an antenna-pattern weighted combination of data from detectors that has the GW signal strain eliminated from it for the correct sky position of the source. Here we propose a network-wide statistic for CBC searches that combines both. With signal and noise simulations we demonstrate that this statistic has the potential for being useful in CBC searches in LIGO and Virgo data.

Our paper is laid out as follows: In Sec.~\ref{preliminaries} we introduce
the conventions and notations used in this
work. Section~\ref{traditionalchisquared} discusses the traditional
$\chi^2$ test for a single detector, followed by how
the null stream can be used as a discriminator in Sec.~\ref{nullstream}. In
Sec.~\ref{idealnullchisquared} we develop the combined null-stream-$\chi^2$ statistic for the simple case of a
network that has all detectors with identical noise power-spectral density. We generalize this statistic to more realistic networks in Sec.~\ref{fullnullchisquared}. Testing of our new statistic in simulated data is presented in
Sec.~\ref{numericaltesting}.

\section{Preliminaries} \label{preliminaries}

Here we describe the waveforms we use to model the CBC signals and the
noise transients or glitches. We also present the matched-filter
based statistic that is at the heart of the CBC searches in GW
detectors.

\subsection{The signal} 
\label{cbcsignal}

We consider non-spinning CBC signals in Advanced LIGO (aLIGO) and Advanced
Virgo (AdV) like detectors in this work. 
The GW strain $h(t)$ in such detectors due to a CBC signal can be expressed in terms of the
plus and cross polarization components $h_{+,\times}(t)$ and the
corresponding antenna patterns $F_{+,\times}$ as follows:
\be
\label{strain}
h(t) = F_+ h_+(t) + F_\times h_\times(t)\,,
\ee
where 
\bea
\label{polorizestrain}
h_+(t) &=& H_+ (m_1, m_2, \epsilon, r) \cos\left[\Psi(m_1, m_2,f_s,t_c,t)\right] \,,\\
h_{\times}(t) &=& H_\times (m_1, m_2, \epsilon, r) \sin\left[\Psi(m_1, m_2,f_s,t_c,t)\right]\,.
\eea
In the above expressions $r$
is the luminosity distance to the binary,
$\epsilon$ is the inclination of the binary's orbit relative to the
line of sight, and the signal phase $\Psi$ depends on the component
masses $m_1$ and $m_2$, apart from auxiliary parameters such as the
detector's seismic cut-off frequency $f_s$, the signal's time of
coalescence $t_c$. Moreover, $H_{+, \times}$ are the two polarization amplitudes.

\subsection{Matched Filtering}
\label{matchedfiltering}

The basic construct used to search for CBC signals is matched
filtering, which involves cross-correlating detector strain data with
templates that are modeled after the waveforms described
above. Detector data comprises noise, $n(t)$, and sometimes a GW
strain signal, $h(t)$. Matched filtering process takes data, $s(t) =
h(t) + n(t)$, and compares it to a template, $Q(t)$, designed to match
the GW strain. It is common to use a normalized complex template
modelled after theoretical waveforms:
\be
\label{unitcomplexsignal}
Q_{(\alpha)}(t) = \mathcal{N}^Q_{(\alpha)}(m_1,m_2,f_s,t_c)\exp \left[i\Psi(m_1,m_2,f_s,t_c,t)\right]\,, 
\ee 
where $\alpha$ is the detector index and $\mathcal{N}^Q_{(\alpha)}$ 
is a (detector noise power spectral density dependent) normalization factor such that $\left<\tilde{Q}_{(\alpha)},\tilde{Q}_{(\alpha)}\right>_{(\alpha)} = 2$. The angular brackets denote an inner product, defined by
\be
\label{crosscorrelation}
\left<a,b\right>_{(\alpha)} = 2 \Re  \int_{0}^{\infty} \frac{\tilde{a}^*(f)
  \tilde{b}(f)}{S_{h (\alpha)} (f)} df\,,
\ee 
where the tilde above a symbol denotes its Fourier
transform and $S_{h (\alpha)} (f)$ is the $\alpha$th detector's two-sided noise power
spectral density (PSD): $\overline{\tilde{n}^*(f)\tilde{n}(f')} =
\delta(f-f')S_h (f)$, with the overbar symbolizing an
average over multiple noise realizations. 
It is this inner product that will be the basis for our statistical
analysis of GW signals. The matched filtering process is performed by
using this inner product between data $s(t)$ and complex templates
$Q(t)$, computed for various values of the signal parameters (such as
$m_{1,2}$). 
Note that the choice of normalization for $\tilde{Q}_{(\alpha)}$ is consistent with our convention where both its real and imaginary parts are each normalized to unity, but 
is at variance with  Ref.~\cite{Allen:2004gu}. Nonetheless such a choice has no effect on the final result.

The matched-filter output is one of the primary ingredients of a decision statistic that allows one to assess if a feature in the GW data is consistent with a GW signal with a high enough significance. The decision statistic may also depend on other characteristics of the data, such as the traditional chi-square~\cite{Usman:2015kfa,Talukder:2013ioa}. When the decision statistic  crosses a preset threshold value for a feature in the data, we will term that feature a trigger. Such features can be noise or signals and are characterized by the decision statistic (and other auxiliary statistics, such as the traditional chi-square), the properties of the template (e.g., $m_{1,2}$), the time of the trigger, etc.
We next define the inner product between data and complex template
as 
\be{}\label{zdefinition}
z \equiv \left<\tilde{Q},\tilde{s}\right>\,.
\ee 
Then the SNR is just
\be
\label{SNR}
{\rm SNR} = \frac{{z}}{ \sqrt{\left<\tilde{Q},\tilde{Q}\right>}} \,,
\ee 
where the denominator is a normalization factor. We again assume the noise to be Gaussian, stationary, and, in the case of multiple detectors, independent of the noise in other detectors.

\subsection{Sine-Gaussian Glitch} \label{sgglitch}

In this study our limited objective is to improve the separation of false positives, caused by
noise transients, from CBC signals in a space similar to that of the
trigger SNR and trigger $\chi^2$. 
We use the following sine-Gaussian function to model noise transients~\cite{Chatterji:2005thesis,Bose:2016jeo}:
\be
\label{glitchstrain}
u(t) = u_0 \sin{(2 \pi f_0 t)} e^{\left(\frac{2 \pi f_0}{K}\right)^2 t^2}\, ,
\ee 
which has amplitude $u_0$ and quality factor $K$. The glitch is centered at frequency $f_0$.
If the amplitude is large and the central frequency lies within the band of a CBC template, then their matched filter can return a SNR value large enough to create a false trigger.

\section{Traditional $\chi^2$  Discriminator} 
\label{traditionalchisquared}

Noise transients can sometimes masquerade as signals and,
thereby, show up as potential detections during matched filtering of
GW detector data with CBC templates.
To combat this issue in a single wide-band detector, such as LIGO
Hanford (H), LIGO Livingston (L), or Virgo (V), the traditional 
$\chi^2$ discriminator was designed to
improve the ability to distinguish between such noise transients and
CBC signals, especially, when the SNR of the noise triggers is
sufficiently high. Let us assume that 
the  noise in a given detector is
Gaussian, stationary, and uncorrelated with that in the other
detectors. 
Next one breaks up the matched filtering integral $z = \left<\tilde{Q},\tilde{s}\right>$ into $p$ smaller sub-bands, where each sub-band is integrated over the frequency interval $\Delta f_j$. Consequently, the matched-filter output can be expressed as
\be
\label{zfreqpartition}
\left<Q,s\right> =2 \Re  \int_0^{\infty} \frac{\tilde{Q}^*(f) \tilde{s}(f)}{S_h (f)} df =2 \Re \sum_{j=1}^p \int_{\Delta f_j} \frac{\tilde{Q}^*(f) \tilde{s}(f)}{S_h (f)} df = \sum_{j=1}^p \left<\tilde{Q},\tilde{s}\right>_j  \,,
\ee 
and one can define $z_j = \left<\tilde{Q},\tilde{s}\right>_j$ to be the matched filtered output over the range $\Delta f_j = [f_{j-1},f_j]$. The frequency partitions can be unequal in frequency range, meaning $\Delta f_1$ can be different in length than $\Delta f_p$. To handle this difference in size in the final statistic we require that the frequency spacing adheres to
\be
\label{qj}
q_j = \frac{1}{2} \left<\tilde{Q},\tilde{Q}\right>_j  \,,
\ee so that the normalization of $\tilde{Q}(f)$ ensures that the sum of the $q_j$'s is unity. 

The usefulness of statistics with $\chi^2$ distributions partly arises from the property that their mean is equal to their number of degrees of freedom. Using the partitioned matched filtering band from above we can design such a statistic by comparing the smaller frequency sub-bands to the total matched filtering output. This is done by taking a difference of the sub-bands with a weighted value of the total
\be
\label{deltaz}
\Delta z_j = z_j - q_j z  \,, 
\ee so that the sum of $\Delta z_j$ over $j$ equals zero. Since these
integrals return complex values, it is important that we take the
modulus squared so that the statistic lies in the reals. The expected
values of these objects lead to the statistic
defined as~\cite{Allen:2004gu}
\be
\label{traditionalstatistic}
\chi^2 =  \sum_{j=1}^p \frac{\Delta z_j^2}{q_j} \,, 
\ee  
where the denominator comes from weighting to ensure that
$\overline{\chi^2} = 2(p-1)$. The statistic is now complete;
and Eq.~(\ref{deltaz}) implies that its mean equals the
number of degrees of freedom. Using Eq.~(\ref{traditionalstatistic}) now distinguishes the glitches from the true signals by difference in the value of this statistic. Glitches will typically return significantly higher values of $\chi^2$ compared to GW signals, for large enough SNRs.

\section{The null-Stream Veto} \label{nullstream}

In addition to single-detector $\chi^2$ discriminators, such as the traditional
one, and multi-detector $\chi^2$ discriminators, the null-stream
construction has also been used to distinguish CBC signals (or parts
thereof) from noise transients~\cite{Colaboration:2011np,Allen:2004gu,Pai:2000zt,Bose:2011km,Harry:2010fr,Babak:2012zx,Talukder:2013ioa,Bose:2016jeo,Dhurandhar:2017aan}.
The null stream is constructed by creating a weighted linear
combination of different detector data streams $s(t)$ in such a way
that contribution from $h(t)$ is eliminated in that combination and all that remains is noise. We may write Eq.~(\ref{strain}) for $D$ number of detectors in a network in the way
\be
\label{vectordata}
\begin{bmatrix}
    \tilde{s}_1  \\
    \tilde{s}_2  \\
    \vdots \\
    \tilde{s}_{D}
\end{bmatrix} = \begin{bmatrix}
    F^+_1 & F^{\times}_1 \\
    F^+_2 & F^{\times}_2 \\
    \vdots & \vdots \\
    F^+_{D} & F^{\times}_{D}
\end{bmatrix}
\begin{bmatrix}
    \tilde{h}_+ \\
    \tilde{h}_{\times}
\end{bmatrix}
+
\begin{bmatrix}
    \tilde{n}_1  \\
    \tilde{n}_2  \\
    \vdots \\
    \tilde{n}_{D}
\end{bmatrix} \, ,
\ee where the data and noise are column vectors, while the strain is
formed from antenna-pattern functions matrix with the GW strain
vector. The detector index then takes the values $\alpha \in [1...D]$. For the following discussion, we define $\textbf{F}^+$ to be a
$D$-dimensional vector with ordered components
$F^+_1,F^+_2,...,F^+_D$, and similarly for $\textbf{F}^\times$.

In the simple hypothetical case of three aligned detectors, with
identical noise PSDs, the null stream~\cite{Chatterji:2006nh} is just
\be
\label{simplenullstream}
N(t) = s_1(t) + s_2(t+\tau_2) - 2 s_3(t+\tau_3)\,,
\ee
which accounts for the correct signal time delays $\tau_{(\alpha)}$ relative to the first (reference) detector to insure that the signal strain cancels out, leaving behind just noise.
A more interesting expression for such a stream can be formed for the
case of non-aligned detectors:
\be
\label{3detectornullstream}
N(t) = A_1 s_1(t) + A_2 s_2(t+\tau_2) + A_3 s_3(t+\tau_3)\,,
\ee
where the components of $\textbf{A}$ can be found from the normalized
cross product $\textbf{F}^+ \times \textbf{F}^{\times}/ ||\textbf{F}^+
\times \textbf{F}^{\times}|| $. For a network with $D>3$, $\textbf{A}$
takes a more general form in terms of $\textbf{F}^{+,\times}$, as
given in Ref.~\cite{Chatterji:2006nh}. 
 In the case of detector data containing signal embedded in noise,
 this linear combination of data vectors $N(t)$ for the right source sky
 position would contain only noise.

The time series of ${\sf n}$ data points that makes up the $\alpha^{\rm th}$
detector's data stream, $s_{\alpha}(t)$, may be thought of as an 
${\sf n}$-dimensional vector. From this we see that the right-hand side of
Eq.~(\ref{vectordata}) is made up of $D$ row vectors, each of which
resides in an ${\sf n}$-dimensional vector space. 
If the detectors are non-aligned, then these $D$ vectors will occupy a $D$ dimensional space. Gravitational wave signals will invariably occupy a two-dimensional sub-space, which is spanned by $\textbf{F}^{+}$ and $\textbf{F}^{\times}$. The remaining $D-2$ basis vectors are then orthogonal to them and define their null space.
We group these $D-2$ basis vectors into a matrix
$\textbf{A}$~\cite{Chatterji:2006nh} with elements:
\be
A_{P\alpha} = \begin{bmatrix}
    A_{11} & A_{12} & \dots & A_{1 D} \\
    A_{21} & A_{22} & \dots & A_{2 D} \\
    \vdots & \vdots & \vdots & \vdots \\
    A_{(D-2)1} & A_{(D-2)1} & \dots & A_{(D-2)D}
\end{bmatrix} \, ,
\ee
where the capital Roman index on $A_{L\alpha}$ is the
null-stream index (ranging from 1 to $D-2$) and the 
Greek letter denotes the detector index (ranging from 1 to $D$).
Each row is normalized to be of unit magnitude, so that
$\sum_{\alpha=1}^D A_{L \alpha} A_{\alpha R}= \delta_{LR}$. If we apply $\textbf{A}$ to Eq.~(\ref{vectordata}) we create a linear combination of data streams that only consists of weighted noise from each detector and no signal. This is termed as the null-stream and arises because the projection of GW signals into the aforementioned $D-2$ dimensional space is zero.

The null-stream  can be used to devise statistics that can discriminate between GW signals
and noise glitches. In Ref.~\cite{Chatterji:2006nh}, the authors construct a whitened null-stream to design a statistical test that compares the energy through cross-correlation and auto-correlation of the data in the discrete domain. In doing so they are able to distinguish between GW burst signals and noise glitches. The test is done by comparing the null energy
\be
\label{nullenergy}
E_{null} = 2 \Re \int_0^{\infty} \tilde{N}^2(f) df\,,
\ee
to
\be
\label{incoherentenergy}
E_{inc} = 2 \Re \int_0^{\infty} \sum_{\alpha = 1}^D B_{\alpha \alpha} \tilde{s}_{\alpha}^2(f) df\,,
\ee
the incoherent energy, with $B_{\alpha \beta} \equiv \sum_{L = 1}^{(D-2)} A_{\alpha L} A_{L \beta} $. 
Whereas the null energy is dependent on both auto-correlation and
cross-correlation of data streams of the various detectors in the
network, the incoherent energy depends only on the auto-correlation
terms~\cite{Chatterji:2006nh}. 
They then use the energy values to distinguish between event types. If a GW signal is present in the data its contribution to the null energy will be eliminated by its cancellation from the null stream construction. When this is the case the relationship between energies is $\overline{E_{null}} < \overline{E_{inc}} $. If there is only noise and glitches in the data then $\overline{E_{null}} \approx \overline{E_{inc}} $. By taking the ensemble average of different noise realizations for the two types of energies they are able to separate the different event types.  As we will show further in our work, this is not the only way null stream construction can be used to create a statistic that is useful in distinguishing between true signal and glitch. 

\section{Idealized Null-Stream-$\chi^2$} \label{idealnullchisquared}

We now wish to design a statistic that incorporates the strengths of
both the traditional $\chi^2$ test as well as null stream veto. This
test would focus on using the matched filtering process to compare
signal in smaller frequency sub-bands, but instead of filtering the
data from a single detector would use the null stream constructed data
from multiple detectors in a network. Done correctly the statistic
will return significantly smaller values for networks that have GW
signal than those containing glitches in their data.

We initially consider a simple case in which one has a network of
unaligned detectors that all have the same noise PSD. We generalize
this to non-identical detector PSDs later. For now we focus on one
null stream, which (after suppressing the null-stream index on
$A_{L\alpha}$) takes the form
\be
\label{detectornullstream}
N = A_1 s_1 (t) + A_2 s_2 (t+\tau_2) +...+ A_{D} s_{D} (t+\tau_{D})\, ,
\ee so that in the case of GW signal the weights $A_{\alpha}$ cancel out all but noise in the data. The goal is to manipulate Eq.~(\ref{detectornullstream}) into a form that takes advantage of the matched filtering. To this end, we multiply both sides of the null stream by the filtering function $\tilde{Q}(f)/S_h(f)$. Renaming the object and integrating gives
\be
\label{singlePSDM}
M = A_1 z_1 + A_2 z_2 +...+A_{D} z_{D}\, ,
\ee where we have taken advantage of all detectors having the same PSD to create the matched filtering outputs $z_{\alpha}$.

It is important to understand the statistical properties of
Eq.~(\ref{singlePSDM}) so that we may use them to mould the final statistic. 
We first look at the average of the square of $M$. This is simplified by knowing
\be
\label{zzaverage}
\overline{z_{\alpha} z_{\beta}} = \left<\tilde{Q},\tilde{h}_{\alpha}\right> \left<\tilde{Q},\tilde{h}_{\beta}\right> + \delta_{\alpha \beta} \left<\tilde{Q},\tilde{Q}\right>\, ,
\ee 
where the Greek letters denote detector index. Having this average makes the work of finding $\overline{M^2}$ simple, with the only unseen complexity coming from the null stream construction. Due to the same denominator in all terms of the integral the null stream constructively cancels cross terms arising from the squaring process, so that
\be
\label{Maverage}
\overline{M^2} = \sum_{\alpha=1}^D A_\alpha^2 \left<\tilde{Q},\tilde{Q}\right> = 2 \,,
\ee due to the properties of $\textbf{A}$ and normalization of the template.

Using the concepts of Allen~\cite{Allen:2004gu}
we now construct 
\be{}\label{deltaMj}
\Delta M_j = M_j - q_j M\,,
\ee 
which is the difference between the contribution to $M$ arising from
the $j^{th}$ sub-band (from a total of $p$ sub-bands) and a weighted
total, 
much like how $\Delta z_j$ was constructed in
Eq.~(\ref{deltaz}). Finding the statistical properties pieces at a
time, without exact specification of frequency bands, we see
Eq.~(\ref{zzaverage}) can be split into $p$ pieces for the $k$th sub-band
\be
\label{zzjaverage}
\overline{z_{\alpha j} z_{\beta k}} = \left<\tilde{Q},\tilde{h}_{\alpha}\right>_j \left<\tilde{Q},\tilde{h}_{\beta}\right>_k + \delta_{\alpha \beta} \delta_{jk} \left<\tilde{Q},\tilde{Q}\right>_j\, ,
\ee so that $j$ is frequency sub-band index. (Note that there is no sum over $j$ in the last term on the right-hand side above.) From Eq.~(\ref{Maverage}) it is manifest that
\be
\label{Mjaverage}
\overline{M_j^2} = \sum_{\alpha =1}^D A_\alpha^2 \left<\tilde{Q},\tilde{Q}\right>_j = \left<\tilde{Q},\tilde{Q}\right>_j \,,
\ee 
which implies that the average is still dependent on $j$. Again, since we have assumed the noise PSD is the same for all detectors we may define the $q_j$ terms as well as the frequency sub-bands from the same integral as in Eq.~(\ref{qj}).

We must now put these terms together to find the properties of
$\overline{M_j^2}$. It will also be important to know $\overline{M_j
  M}$. As Allen
does~\cite{Allen:2004gu} we will use symmetry, and the fact that
\be
\label{symmetryM}
\sum_{j=1}^p M_j M = M^2
\ee to find that both cross terms of $\overline{\Delta M_j^2}$ are $q_j \overline{M_j^2}$. Combining these properties leads to
\bea
\label{a}
\overline{\Delta M_j^2} &=& \overline{(M_j - q_j M)^2} \\
&=& \overline{M_j^2} -q_j \overline{(M_j M)^2} -q_j \overline{(M_j M)^2} +q_j^2 \overline{M^2}\\
&=& \left<\tilde{Q},\tilde{Q}\right>_j - 2 q_j^2 \overline{M^2} + q_j^2 \overline{M^2} \\
&=& 2q_j (1 - q_j)\, ,
\eea dependent only on the weights defined by the number of frequency
sub-bands chosen. Much like Allen~\cite{Allen:2004gu}
we take this result and define our statistic
\be
\label{rho1}
\rho = \sum_{j=1}^p \frac{\Delta M_j^2}{q_j} \,,
\ee giving it an average of $2(p-1)$. Thus $\rho$ is $\chi^2$
distributed with its average being the same as the degrees of
freedom. Lastly, to account for multiple null streams for cases when
the network has four or more detectors we generalize
$\rho$ by including their contributions.
We do so by first reviving the null-stream index in
Eq.~(\ref{singlePSDM}) such that
\be
\label{singlePSDMnullstreamL}
M_{L} \equiv A_{L1} z_1 + A_{L2} z_2 +...+A_{L D} z_{D}\, ,
\ee
and define $\Delta M_{Lj}$ for each null-stream similar to the one in Eq.~(\ref{deltaMj}).
We finally generalize the $\rho$ in Eq.~(\ref{rho1}) by including the sum over
the null-streams to arrive at:
\be
\label{rhoPSDnull}
\rho = \sum_{L=1}^{D-2} \sum_{j=1}^p \frac{\Delta M_{L j}^2}{q_j} \,,
\ee noting that the $q_j$ values are unaffected due to the same noise PSD in all detectors. From Eq.~(\ref{rhoPSDnull}) it is clear to see the degrees of freedom are accounted for in the average, coming out to $2(D-2)(p-1)$.

\section{Full Null-stream-$\chi^2$} \label{fullnullchisquared}

Having explained the basic idea of the null-stream-$\chi^2$ statistic
in Sec.~\ref{idealnullchisquared}, we develop it further here so that
the resulting statistic addresses some of the challenges associated
with real data. One such challenge is that the noise PSD is very
likely to vary from one detector to another. Another one is the fact
that the detectors are oriented differently around the globe.

We begin by pointing out that Eq.~(\ref{Maverage}) was clean and
concise because 
all detectors there were assumed to have the same noise PSD. 
To address the fact that these noise PSDs will be different in
general, 
we first introduce the over-whitened data stream $\tilde{s}_{w\alpha} \equiv \tilde{s}_{\alpha} / S_{h \alpha}$. Similarly, the over-whitened antenna-pattern functions are used to construct the 
$\textbf{F}$ matrix:
\be
\label{whitenedFmatrix}
\begin{bmatrix}
    F^+_1/S_{h1} & F^{\times}_1/S_{h1} \\
    F^+_2/S_{h2} & F^{\times}_2/S_{h2} \\
    \vdots & \vdots \\
    F^+_{\alpha}/S_{h\alpha} & F^{\times}_{\alpha}/S_{h\alpha}
\end{bmatrix}\,,
\ee
where we have divided each $F^{+,\times}_\alpha$ by the noise PSD of the corresponding detector. 
The $A_{w \alpha}$ are obtained from the weighted
antenna-pattern functions the same way the $A_{\alpha}$ are deduced
from the unweighted ones above. In the case of a network with three non-aligned detectors the $A_{w \alpha}$ take the form
\be{}
\label{threedetnetAw}
A_{w \alpha} = \frac{S_{h\alpha} A_{\alpha}}{\sqrt{\sum_{\alpha=1}^3 S_{h\alpha}^2 A_{\alpha}^2 }}\,,
\ee which shows their explicit dependence on the $A_{\alpha}$ and detector noise PSDs.
They are now frequency dependent, owing to the newly incorporated PSD
factors.
We also define the bracket operation to define the following inner product:
\be
\label{newfilter}
\left[Q,b_w\right] = 2 \Re \int_0^{\infty} \tilde{Q}^*(f) \tilde{b}_w(f) df\,,
\ee 
where $\tilde{b}_w(f)$ is obtained by over-whitening $\tilde{b}(f)$.
Therefore, it is clear that $[Q,s_{w \alpha}] = z_{\alpha} $.

Up to this point the templates used have been constructed from two
orthonormal pieces for each polarization. Due to the frequency
dependence in the $A_{w\alpha}$ terms we choose to construct our
statistic filtering for each polarization separately, this motivation
will become clear shortly. To handle the polarizations separately we
note that the complex filter $Q$ can be decomposed into its real and
imaginary parts such that
\be
\label{plusfilter}
\tilde{Q}_+ = \Re(\tilde{Q})
\ee
and
\be
\label{crossfilter}
\tilde{Q}_{\times} = \Im(\tilde{Q})\,,
\ee
where we will use a network-wide plus and cross filter on the data. We will focus on the plus polarization first:
\be
\label{networkPSDM}
W_+ =  [Q_+,N_w]\, ,
\ee where similar to Sec.~\ref{idealnullchisquared} we constructed a
filtered null stream, with $N_w$ being the over-whitened null stream. With this network template we can now focus on the matched-filter output outlined in Eq.~(\ref{newfilter}) to be the basis for constructing a network-wide statistic that will be $\chi^2$ dependent.

Proceeding in the same vein as Allen~\cite{Allen:2004gu}, we are interested in the mean of the square of Eq.~(\ref{networkPSDM}). Understanding the mean of $W_+^2$ can be aided by defining
\be{}\label{lambdameanfunction}
\Lambda \equiv \sum_{\alpha}^{D} A_{w \alpha}^2/S_{h \alpha}\,,
\ee motivated by accounting for the possible differences in detector noise PSDs as well as the construction of the $A_{w\alpha}$ given in Eq.~(\ref{threedetnetAw}). The mean takes the form,
\be
\label{whiteMaverage}
\overline{W_+^2} =  [Q_+, \Lambda Q_+ ]\, ,
\ee where the null stream construction $N_w$ has any contribution from a GW signal removed. We use this result to complete our construction of the statistic. This is done by renormalizing our detector templates so that $[Q_+,\Lambda Q_+]=[Q_{\times},\Lambda Q_{\times} ]=1$. For the mean of $W_+^2$ we then simply have one.

Turning our attention to breaking the frequency space into smaller bands we define the bands by
\be
\label{geoqj}
q_{+j} = [Q_+, \Lambda Q_+ ]_j \, ,
\ee similarly to the previous section. This insures that our partitions
$q_j$ also sum to one. With our new constructs we may now build our
new statistic. As Ref.~\cite{Allen:2004gu} does, we take a difference so that $\Delta W_+ = W_{+j} - q_{+j} W_+$. Doing so leads to a statistic much like Eq.~(\ref{rhoPSDnull})
\be
\label{plusrhonull}
\rho_{+} =  \sum_{j=1}^p \frac{\Delta W_{+ j}^2}{q_{+j}} \,,
\ee that is $\chi^2$ distributed with a mean of $p-1$. While this statistic has the desired properties, it only filters for the plus polarization. A similar statistic can be constructed for the cross polarization, the only difference will be to use $Q_{\times}$ to construct $W_{\times}$. In doing so we have
\be
\label{crossrhonull}
\rho_{\times} =  \sum_{j=1}^p \frac{\Delta W_{\times j}^2}{q_{\times j}} \,,
\ee
that is also $\chi^2$ dependent with a mean of $p-1$. To complete our
statistic we note that the addition of two $\chi^2$ statistics is also
$\chi^2$ with a mean equal to the sum of the means of the original
statistics. We then combine the cross and plus matched filter outputs
to create the complete statistic:
\be
\label{fullrhonull}
\rho_{f} = \sum_{L=1}^{D-2} \sum_{j=1}^p \left[ \frac{\Delta W_{+\alpha j}^2}{q_{+j}} + \frac{\Delta W_{\times \alpha j}^2}{q_{\times j}}\right] \,,
\ee
where we have been careful to sum over the different null streams dependent on the number of detectors in the network. We see that $\rho_f$ is $\chi^2$ dependent with a mean of $2(D-2)(p-1)$, and constructed using a null stream focus on matched filtered outputs from a network of detectors.

This statistic takes full advantage of all the detectors in the
network even when they have differing noise PSDs. The null stream
construction allows for the elimination of the signal strain from our
statistic when the correct sky location of the source is used. 
As we show below, $\rho_f$ can be utilized in the construction of a decision statistic (along with the SNR) that compares favorably with other alternative statistics in discriminating signal triggers from a class of noise-transient triggers.

\section{Numerical Testing of The Null-stream-$\chi^2$ Discriminator Statistic} 
\label{numericaltesting}

In this section we describe the signal and noise artifact simulations
conducted to test the discriminatory power of the new statistic
defined by Eq.~(\ref{fullrhonull}). 
For modeling noise transients we limit ourselves to sine-Gaussians, which
have been shown to be a good (but not necessarily complete) basis for
modeling glitches in real detector data~\cite{Chatterji:2005thesis,Bose:2016jeo,Bose:2016sqv,Nitz:2017lco}.
We do this by examining how the traditional $\chi^2$ test in
Eq.~(\ref{traditionalstatistic}) performs in separating distributions
of signal and noise triggers:
If a true GW signal is in the data then a band by band comparison in
the frequency domain of the anticipated signal power and the actual power in
the data, is expected to yield smaller $\chi^2$ values than when there
is a noise transient instead.
In our tests on simulated data we also explore the usefulness
of increasing the number of degrees of freedom by partitioning the
signal band into more sub-bands in computing the null-stream
statistic.
This facility in our null-stream construction may help in improving
the discriminatory power of the null-stream statistics discussed by Chatterji \textit{et al}.~\cite{Chatterji:2006nh} and Harry \textit{et al}.~\cite{Harry:2010fr}. 

\begin{figure}
\centerline{\includegraphics[scale=.35]{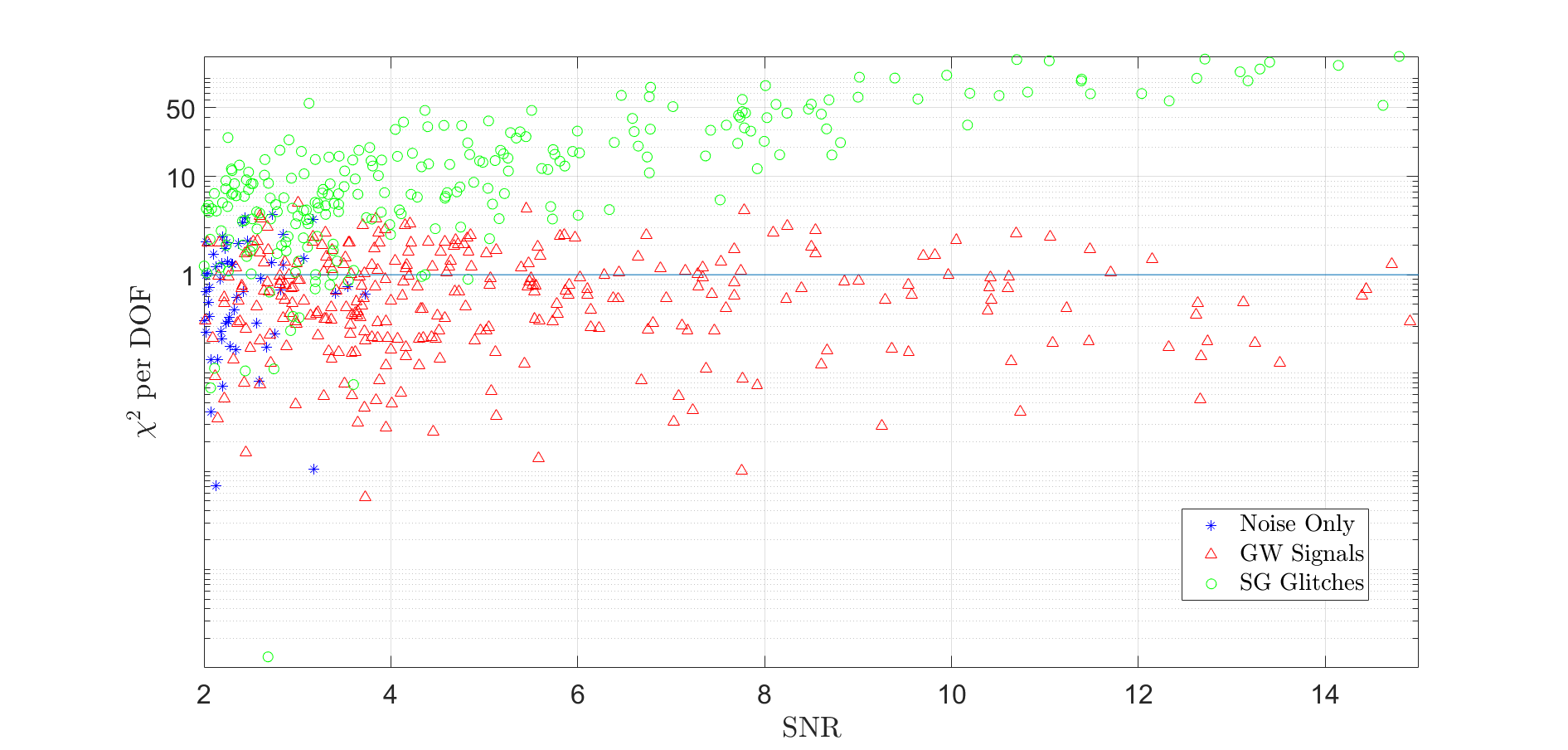}}
\caption{
The traditional $\chi^2$, Eq.~(\ref{traditionalstatistic}), per
degree of freedom (DOF) plotted against the signal-to-noise ratio
(SNR)\, Eq.~(\ref{SNR}), for various kinds of triggers in our simulation studies, for $p=12$
$\chi^2$ sub-bands, in a single detector. The
triggers were generated for Gaussian noise alone (blue stars), binary black
hole signals (red triangles), and sine-Gaussian glitches (green
circles). Gaussian noise triggers are expected to have an average $\chi^2$
per DOF of unity. The signal triggers have a similar average above,
for various signal strengths (and SNRs). With increasing SNR, the
$\chi^2$ distribution of glitches separates more and more from that
of signals. (Here we took the signal
parameters to match the template parameters exactly. A parameter
mismatch will cause the signal trigger $\chi^2$ to rise with
increasing SNR.)
}  
\label{fig:AllenNumericalSim}
\end{figure}

The BBH signals we simulated were based on the  (frequency-domain) IMRPhenomD model~\cite{Husa:2015iqa,Khan:2015jqa}.~\footnote{Since the detector data is in the time domain, ideally one should simulate signals and add them to noise (simulated or real) in the same domain, even if the matched-filtering is implemented in the frequency domain. We aim to carry out such studies in the future.} The components were non-spinning and had masses chosen from the range
$(10,30)~M_\odot$. The BBHs were distributed uniformly in volume between a luminosity distance of 1~Gpc and 3~Gpc. 
All noise transients were simulated to
be sine-Gaussians (SGs), as introduced in Sec.~\ref{preliminaries}, with quality factor $K\in (10,45)$ and central frequency
$f_0 \in (80, 200)$~Hz. 
The strength of 
the SG glitches was taken to be such
that the single detector matched-filter SNR was below $\approx 20$. 
We first present our results for the traditional $\chi^2$ test, given
in Eq.~(\ref{traditionalstatistic}) (Ref.~\cite{Allen:2004gu}), in
Fig.~\ref{fig:AllenNumericalSim}, where the
reduced $\chi^2$ (i.e., $\chi^2$ per degree of freedom) values are plotted versus the SNR for three different types of
triggers, namely, Gaussian noise, BBH signals, and SG glitches. 
As is expected of Gaussian noise and signals embedded in such noise,
their $\chi^2$ per degree of freedom (DOF) values distribute with
average around one. The SG glitch triggers carry higher values of $\chi^2$ with
increasing SNR, which allows this test
to distinguish the signal from such noise triggers better at higher SNRs.

Having shown that our simulations yield results along lines expected
of the traditional $\chi^2$ test
~\cite{Allen:2004gu}, we move onto testing the null-stream statistic
in Eq.~(\ref{fullrhonull}) using similar simulated data.
Since we are now modeling a network of detectors we first create the
null stream from Eq.~(\ref{detectornullstream}) for the Hanford,
Livingston, Virgo (HLV) network. For the network test, we limit
ourselves here to the targeted search, where the sky position of the
GW source is known in advance~\cite{Harry:2010fr}, e.g., from the location of a putative
electromagnetic counterpart, such as a short-duration gamma-ray
burst~\cite{Loeb:2016fzn}. (A blind search requires more computational time or resources and may also incur some deterioration in the search performance. We will pursue that study in a subsequent work.)
The sky position information is used to compute the antenna-pattern vectors ${\bf F}_{+,\times}$, the correct
time delays for signals across the detector baselines, as well as the
$A_{\alpha}$ factors
for the null stream construction. 
When studying SG glitches, we consider two cases: (a) 
There is a SG glitch in one of the detectors but only
Gaussian noise in the other two; (b) There are near concurrent SG
glitches, with varied $K$ and $f_0$ values, in two of the detectors, and only
Gaussian noise in the third. When we vary $K$ and $f_0$ values they are chosen from the previously described range of values, but the glitch characteristics are different in the two detectors, such that the difference in $K$ is 5 or more and that in $f_0$ is at least 10Hz.
While the second case is expected to be much rarer than the first one
in real data, it may assume importance in situations when we ascribe
false-alarm rates to our detections at the level of one in several
tens of thousands of years.
For the multi-detector simulations, when comparing the performance of different $\chi^2$ statistics, the same simulated signals, glitches and noise are used.

\begin{figure}
\centerline{\includegraphics[scale=.35]{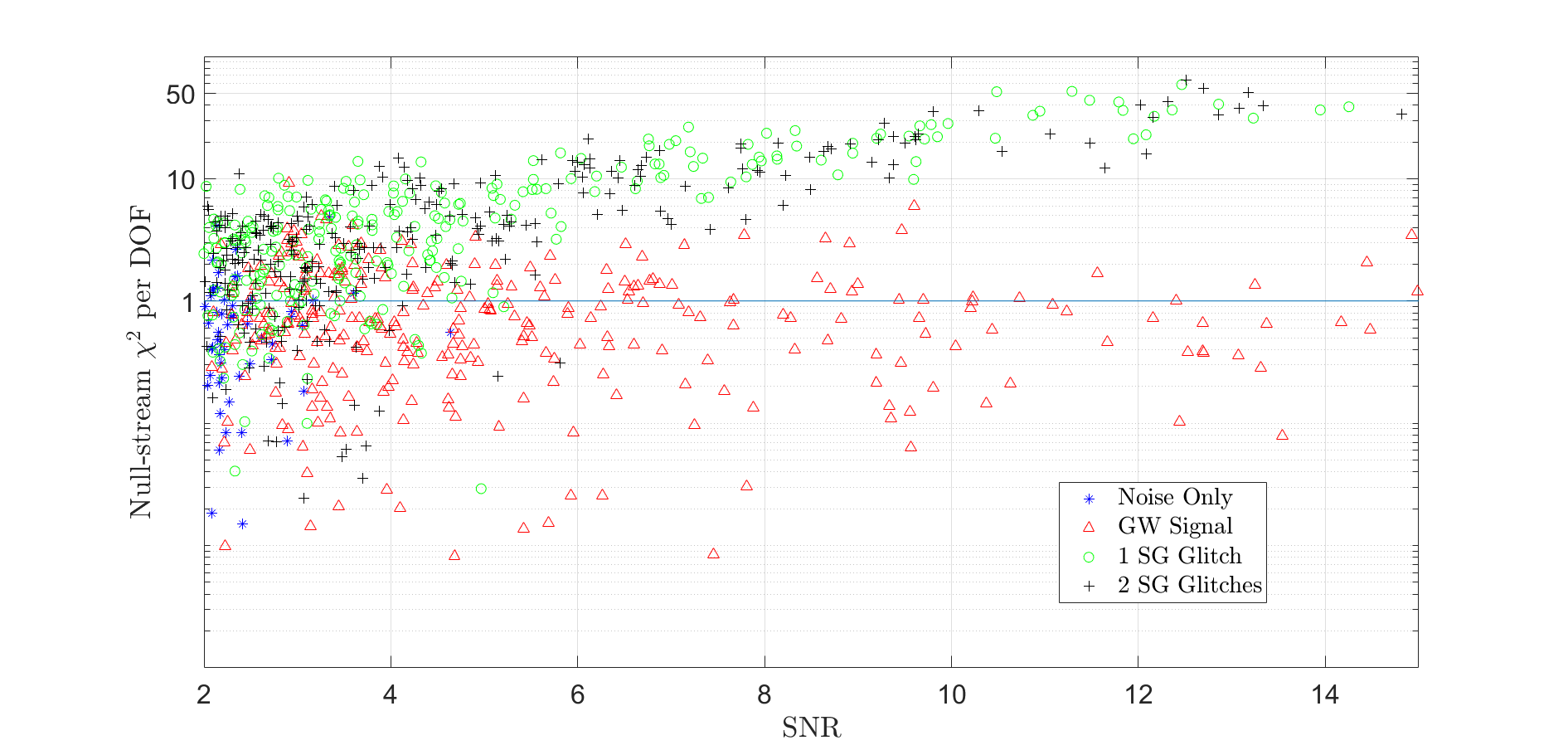}}
\caption{Like in Fig.~\ref{fig:AllenNumericalSim}, here too we compare
  noise only, BBH signal, and SG glitch triggers, but in the HLV network. The other differences
  are: (a) on the vertical axis the null-stream-$\chi^2$ per DOF is plotted for
  $p=2$  sub-bands and (b) in black pluses triggers arising from
  concurrent sine-Gaussian glitches (with different parameters) in two
  of the three detectors have been included.
The SNR on the horizontal axis is the combined SNR in the HLV network (defined following Eq.~(\ref{newstat})).
The null-stream-$\chi^2$ statistic is the same as the one defined in 
Eq.~(\ref{fullrhonull}) for the HLV network.
It shares many features with Fig.~\ref{fig:AllenNumericalSim}, with
noise only and GW signal triggers both having an average approximately
equal to unity. Since the number of sub-bands is small, at $p=2$, the relative
contribution of the null stream as a discriminator, vis \'{a} vis the traditional
$\chi^2$-statistic, is large. 
When $p$ is increased to 12, the traditional
$\chi^2$ aids in improving this discrimination, as shown in
Fig.~\ref{fig:NullStreamNumericalSim_12bin}.
}   
\label{fig:NullStreamNumericalSim_2bin}
\end{figure}

\begin{figure}
\centerline{\includegraphics[scale=.35]{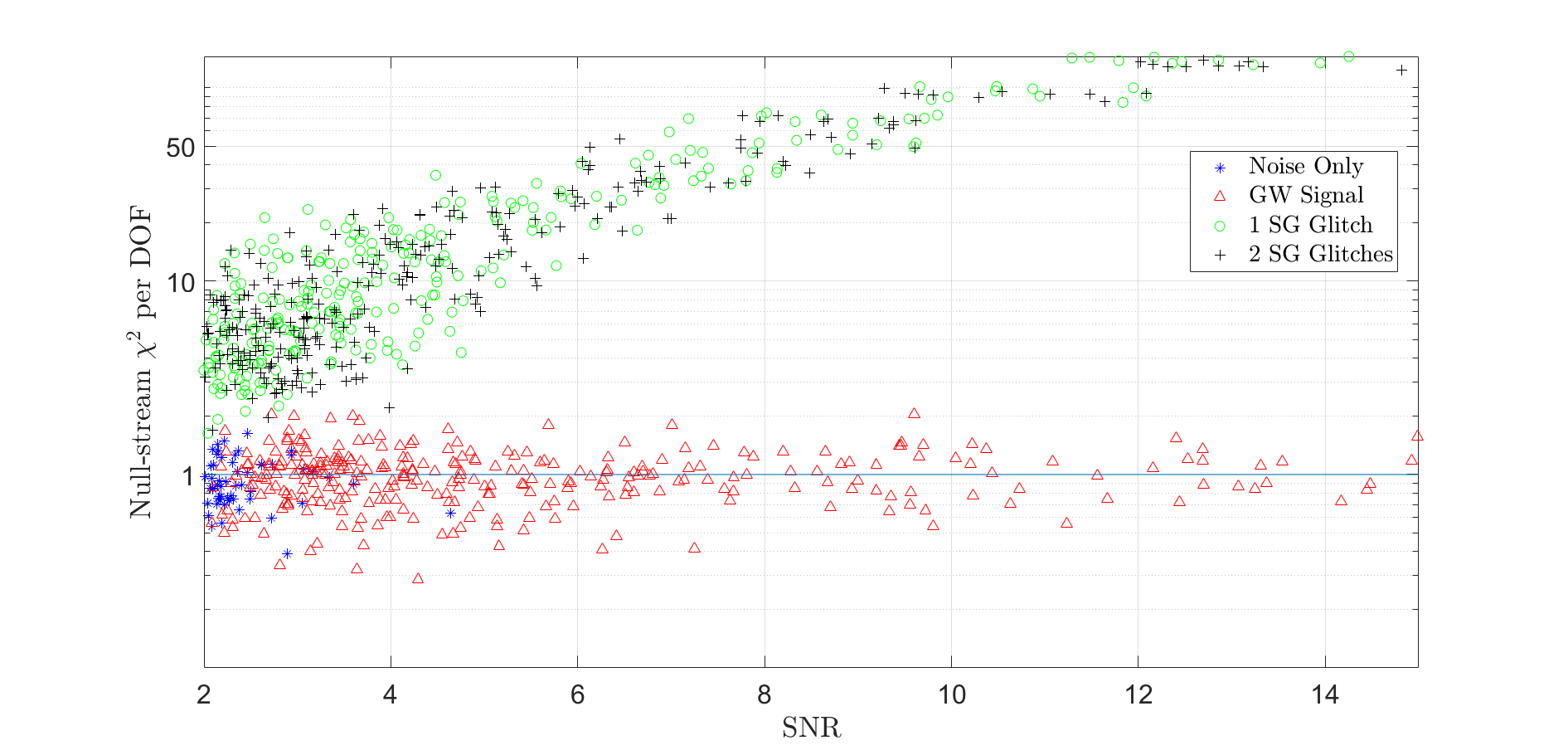}}
\caption{Same as Fig.~\ref{fig:NullStreamNumericalSim_2bin}, but with $p=12$ sub-bands.}
\label{fig:NullStreamNumericalSim_12bin}
\end{figure}

\begin{figure}
\centerline{\includegraphics[scale=.40]{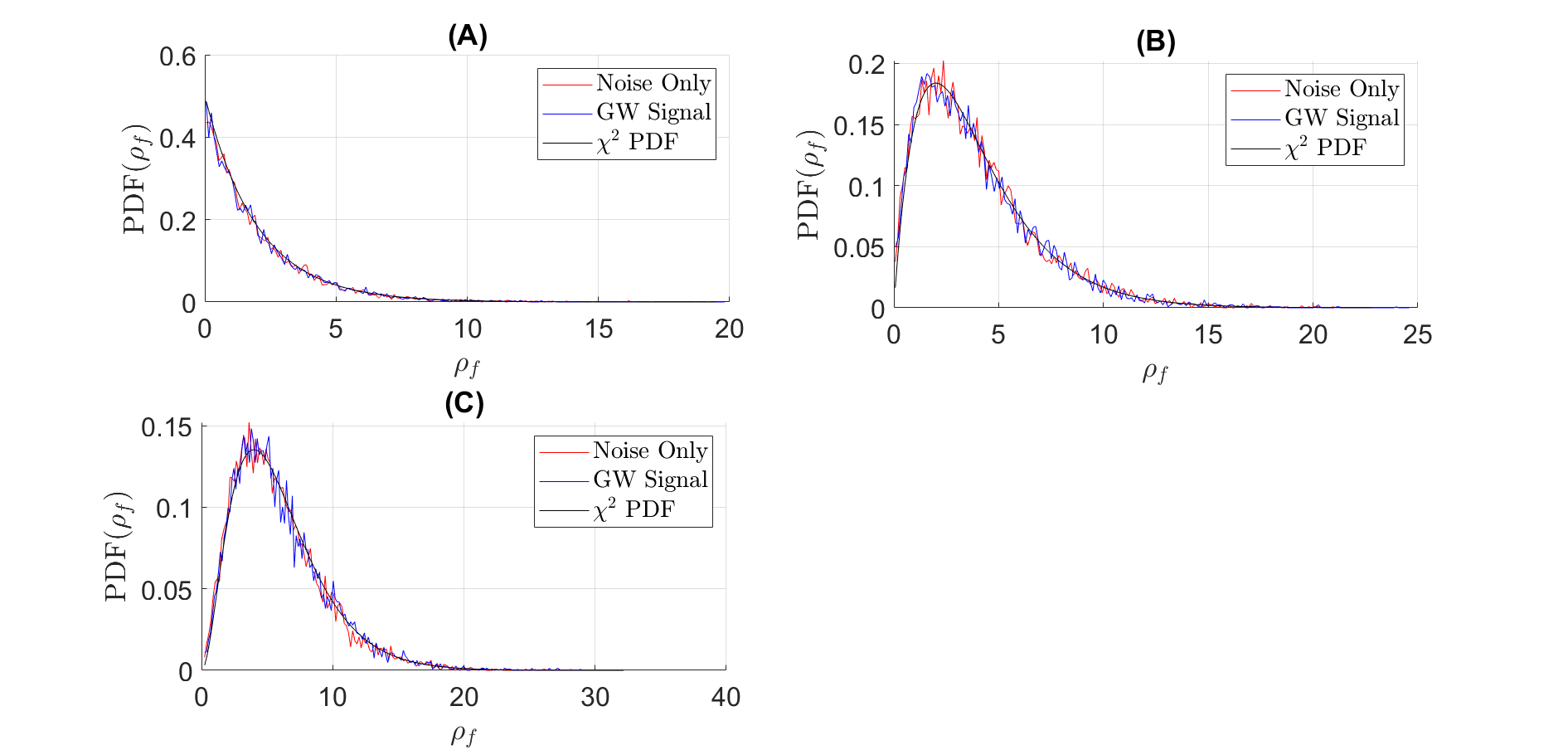}}
\caption{Here we show a histogram comparison of $\rho_f$ values for noise only data and GW signal data for our null-stream-$\chi^2$ test as described by Eq.~(\ref{fullrhonull}) using the same 3 detector method described at the beginning of Sec.~\ref{numericaltesting}. These histograms are compared to the analytic probability density function (PDF) for a $\chi^2$ statistic of a given degree of freedom. We include $p=2$ for (A), $p=3$ in (B), and $p=4$ for (C). These each correspond to a degree of freedom of $2$,$4$, and $6$ respectively. This shows explicitly the $\chi^2$ dependence that $\rho_f$ carries.}   
\label{fig:whitenedPDFcompare}
\end{figure}

\begin{figure}
\centerline{\includegraphics[scale=.40]{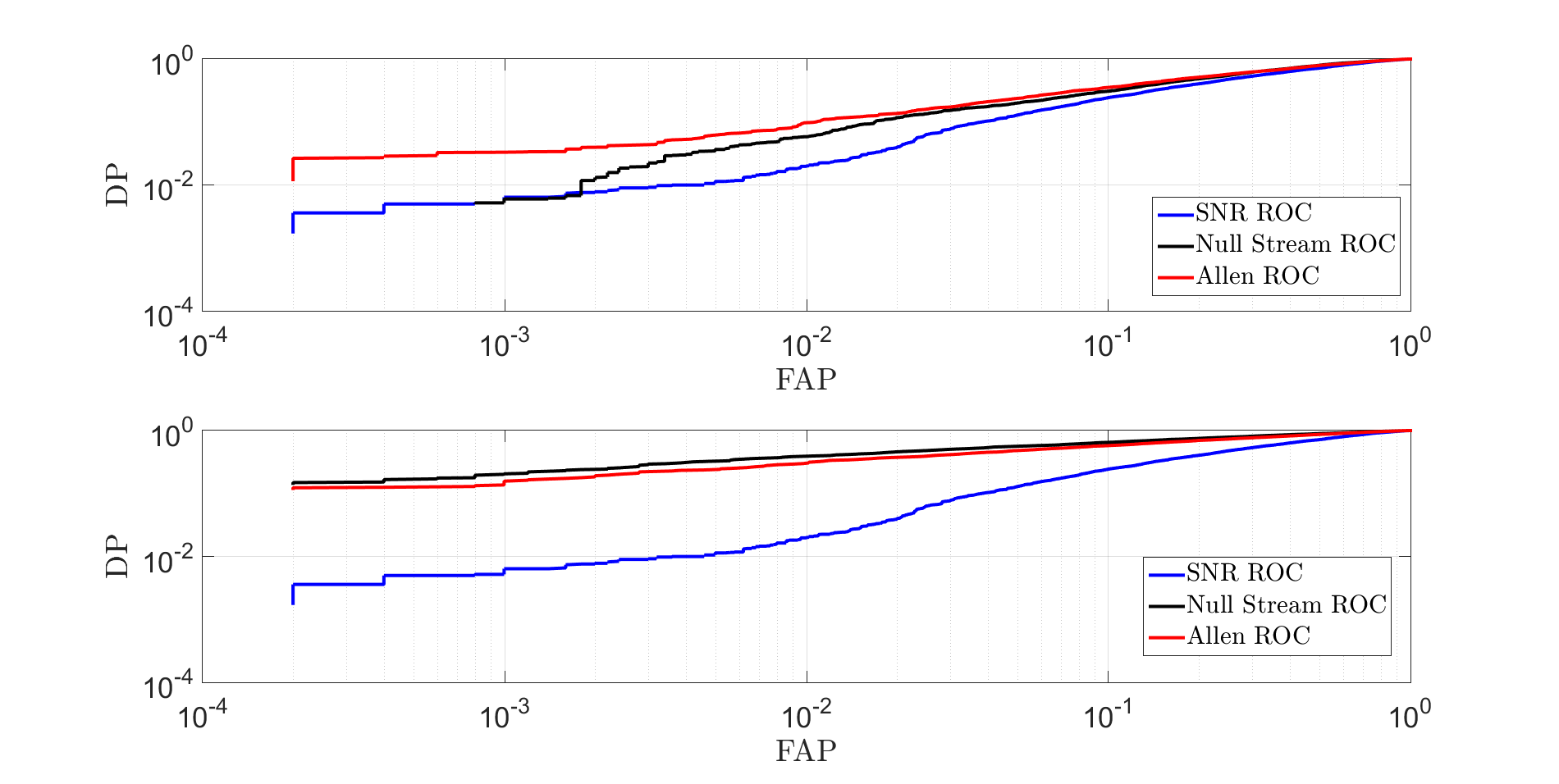}}
\caption{
We compare the performances of the network traditional
$\chi^2$ (red) and the null-stream-$\chi^2$ (black)
discriminators by plotting the receiver operating
characteristics (ROCs) of $\zeta_T$ and $\zeta_N$, respectively, for $p=2$ (top panel) and $p=12$ (bottom panel)
sub-bands. The ROCs for just the combined SNR, without the use of any of the aforementioned $\chi^2$ statistics, are
shown in blue. Increasing the number of sub-bands from $p=2$ to 12 not only improves
the performance of the network traditional $\chi^2$ (which is expected), but also that of the
null-stream-$\chi^2$, but the blue curve, understandably, remains unchanged. While for $p=12$, in our limited study, the null-stream-$\chi^2$ happens to perform slightly 
better than the traditional $\chi^2$ for all FAP values that we were able to
explore, we caution that this performance may further improve or
worsen in real data, and can vary from one type of noise transient to
another.
}   
\label{fig:ROCcomparison}
\end{figure}

For $p = 2$ sub-bands, qualitatively, Fig.~\ref{fig:NullStreamNumericalSim_2bin} shows that
while the null-stream test performs well in discriminating single SG
glitches from BBH signals, it does much worse for double SG glitches.
This is to be expected in this case due to the small number of $\chi^2$ sub-bands
($p=2$) that are paired with the null stream construction. If there is
a second glitch in the network, the null stream may share more
overlap in the frequency bands used and thus not contribute as much to
the $\chi^2$ value. This is where the ability to test the data in
a larger number of sub-bands becomes useful, as seen in
Fig.~\ref{fig:NullStreamNumericalSim_12bin}, which uses $p=12$
sub-bands. By increasing the number of sub-bands one is able to check
for better time-frequency consistency of transient patterns in multiple
detectors. Ideally, the value of $p$ is best chosen by comparing the
performance of the null-stream-$\chi^2$ statistics for different
values of $p$ in real data, with real noise transients. This ``tuning"
problem will be addressed in a subsequent work. The explicit $\chi^2$ distribution dependence of our statistic, $\rho_f$, can be seen in Fig.~\ref{fig:whitenedPDFcompare}.

Finally, for an assessment of the power brought in by the null-stream-$\chi^2$, we constructed two multi-detector statistics, 
\begin{equation}
\label{newstat}
\zeta_{\rm T,N} \equiv \frac{\rm SNR_c}{\left( \chi^2_{\rm T,N}
\right)^{1/3}}\,,
\end{equation}
where ${\rm SNR_c}$ denotes the detector network's {\em combined} signal-to-noise ratio, which is defined as $[\sum_{\alpha=1}^D {\rm SNR}^2_\alpha]^{1/2}$, with ${\rm SNR}_\alpha$ being the signal-to-noise ratio of a trigger in the $\alpha^{\rm th}$ detector, $\chi^2_{\rm T}$ is the {\em network} traditional $\chi^2$ statistic per degree of freedom (DOF) and  $\chi^2_{\rm N}$ is the null-stream-$\chi^2$ statistic ($\rho_f$) per DOF for the network. The network traditional $\chi^2$ statistic is the sum of the traditional $\chi^2$ statistics over all detectors in that network.
Triggers with large values of $\chi^2_{\rm T,N}$ are penalized by the new statistics $\zeta_{\rm T,N}$. Constant $\zeta_{\rm T}$ curves represent constant false-alarm probability (FAP) curves on a $\chi^2_{\rm T}$ {\it versus} ${\rm SNR_c}$ plane in our HLV  simulations. Thus, the fraction of all simulated signals with $\zeta_{\rm T}$ values above a threshold $\zeta_{{\rm T}0}$ denotes the detection probability for this statistic at a constant FAP. 
Constant $\zeta_{\rm N}$ curves, on the other hand, represent approximately constant false-alarm probability (FAP) curves on a $\chi^2_{\rm N}$ {\it versus} ${\rm SNR_c}$ plane for the same HLV simulations. 

We use the $\zeta_{\rm T,N}$ statistics to construct receiver-operating characteristic (ROC) curves~\cite{Helstrom} in Fig.~\ref{fig:ROCcomparison}.
As shown there,
increasing $p$ from 2 to 12 improves the performance of both the
network traditional $\chi^2$ based statistic, $\zeta_{\rm T}$, and the null-stream-$\chi^2$ based one, $\zeta_{\rm N}$. The degree
of improvement is more for the latter, but the performance of both
statistics is comparable for  $p=12$. This holds out hope that the 
null-stream-$\chi^2$ can be developed further to improve its ability
to discriminate noise transients from signals (at least in  certain
sections of the signal parameter space) in real data. 
This proposition assumes significance now, given that in
addition to the existing LIGO and Virgo detectors, KAGRA (Japan) and
LIGO-India are being constructed, and it is likely that joint
multi-detector analysis with three or more detectors will be pursued in that network, which will allow the application of statistics such as the null-stream-$\chi^2$.

Further study and possible extensions of the null-stream-$\chi^2$ will
involve, e.g., finding the optimal
value of $p$, identifying useful regions in the signal parameter space for
its implementation, etc. Since these factors depend on the nature of
the glitches, we plan to carry out such studies in real data, beginning with
LIGO and Virgo, in the future.

The unified construction of $\chi^2$ tests~\cite{Dhurandhar:2017aan} has shown that some of the existing discriminators, such as the traditional $\chi^2$ studied here, target a relatively small part of the space occupied by detector data. Therefore, it is imaginable that as the detectors become more sensitive new noise transients may arise that do not lie in that subspace and yet provide good overlap with CBC templates. The null-stream-$\chi^2$ may be useful against such artifacts. Moreover, the fact that the effectiveness of the null stream itself is  less impacted by  mismatches between the CBC template model and the signal in the data can also prove useful in devising better extensions of the null-stream-$\chi^2$.
However, it remains to be seen how it can be useful in blind searches.
The hope is that the null-stream-$\chi^2$ will
complement the existing discriminators in reducing the significance
of certain glitches in some sections of the parameter space of CBC searches in multi-detector data.

\section{Conclusion} 
\label{conclusion}

In this work we introduced a new multi-detector statistic -- the
null-stream-$\chi^2$ -- that can
be developed further for discriminating between CBC signals and noise
transients in real data. We did so by implementing the traditional
$\chi^2$ test on a noise-PSD weighted null stream.
The new statistic follows a
$\chi^2$ distribution by design. We studied its performance by applying
it to multi-detector simulated data, some subsets of which had simulated BBH
signals and sine-Gaussian glitches added separately. We constructed SNR
vs $\chi^2$ plots and receiver-operating
characteristics~\cite{Helstrom} to demonstrate that the new statistic compares well
with statistics devised in the past in distinguishing signals from
noise transients.

Null stream vetoes can be effective when the exact model or parameters of the
signal are not known, but they require careful construction when detector sensitivity levels vary~\cite{Harry:2010fr,Chatterji:2006nh}. As we showed here, even for modeled signals their extension, in the specific form of the null-stream-$\chi^2$ statistic, has the
potential to be useful in multi-detector CBC searches, at least when
the detectors' sensitivities to the common BBH source (as quantified by their horizon distances to it) are not
very different. This first demonstration, however, has been for targeted searches. Since it is not clear if BBHs have electromagnetic counterparts a targeted search with BBH templates may not appear to be of much use. This point may be debatable~\cite{Loeb:2016fzn}, and some may argue for employing higher mass templates than just binary neutron star ones for targeted searches of gamma-ray bursts. In such an event, a $\zeta_{\rm N}$-like statistic can be useful. A more conservative scenario is one where a promising BBH candidate is found by other statistics in a blind search, and its parameters are then used as in a targeted search by  $\zeta_{\rm N}$ (or its extended version for real data), as a follow-up, to improve the significance of that candidate. The real worth will instead be in demonstrating that $\zeta_{\rm N}$ can be developed further to work in blind searches more directly (i.e., not as a follow up), and explore the limits of its performance when the relative sensitivities of the detectors in the network are varied. This is what we plan to do in future.

\section*{Acknowledgments}

We thank Bhooshan Gadre for carefully reading the paper and making useful comments. This work is supported in part by the Navajbai Ratan Tata Trust and NSF grant PHY-1506497. This paper has the LIGO document number {\color{red} LIGO-DCC-P1800334}.

%\begin{references} 

\end{document}